\documentclass[12pt]{article}

\usepackage{inputenc}
\usepackage[OT1]{fontenc}

\usepackage[nodisplayskipstretch]{setspace} 
\usepackage[margin=1.25in]{geometry}

\usepackage[usenames,dvipsnames,svgnames]{xcolor}
\usepackage{amsmath, amssymb, amsfonts, graphicx, pdflscape, mathtools, amsthm, upgreek, bm,pgfplots,csvsimple,multirow, multicol, booktabs, import, placeins, subcaption, microtype,etoolbox,thmtools}

\usepackage{tikz}
\usetikzlibrary{shapes.geometric, arrows, fit, positioning}
\usepackage[multiple]{footmisc}
\usepackage{verbatim}
\usepackage[nottoc,numbib]{tocbibind}

\usepackage[backend=biber,style=authoryear,sorting=ynt,maxbibnames=9,citetracker=true,maxcitenames=2,uniquename=false,uniquelist=false,natbib]{biblatex}

\usepackage{verbatim}
\usepackage{accents}
\newcommand{\ubar}[1]{\underaccent{\bar}{#1}}
\usepackage{needspace}

\usepackage{comment}
\usepackage{enumitem}
    \setlist[itemize]{noitemsep,nolistsep}
    \setlist[enumerate,1]{noitemsep,nolistsep,label=(\arabic*)}
    \setlist[enumerate,2]{noitemsep,nolistsep,label=(\alph*)}
    \setlist[enumerate,3]{noitemsep,nolistsep,label=(\roman*)}
\usepackage{tocvsec2}
\usepackage{threeparttable}
\usepackage[page,title,toc]{appendix}
\usetikzlibrary{calc, cd}
\setlist{noitemsep, topsep=0pt}
\usepackage[skip = 10pt]{caption}
\allowdisplaybreaks[1]
\allowdisplaybreaks[1]

\makeatletter
\renewcommand{\paragraph}{%
	\@startsection{paragraph}{4}%
	{\z@}{1.5ex \@plus 1ex \@minus .2ex}{-0.7em}%
	{\normalfont\normalsize\bfseries}%
}
\makeatother

\definecolor{Blue}{RGB}{86,180,233}
\definecolor{Orange}{RGB}{230,159,0}
\definecolor{Green}{RGB}{0,158,115}
\definecolor{GmailBlue}{RGB}{42, 93, 176} 
\usepackage[
	colorlinks=true,
	citecolor= GmailBlue,
	linkcolor=GmailBlue,
	urlcolor = GmailBlue,
]{hyperref}

\usepackage{pgfplots}
\usepgfplotslibrary{groupplots,colorbrewer}
\pgfplotsset{compat=newest}
\pgfplotsset{cycle list/Set1}
\usepackage{tikz}
\usetikzlibrary{matrix,calc,shapes,arrows.meta,positioning}
\tikzset{
    vertex/.style = {shape=circle,draw, minimum size = 1.8em, inner sep = 0pt},
    edge/.style = {->,> = latex}
}

\usepackage[capitalize,noabbrev,nameinlink]{cleveref}

\theoremstyle{plain}
  \newtheorem{theorem}{Theorem}
  \newtheorem{lemma}{Lemma}
  
  \newtheorem{corollary}{Corollary}

\theoremstyle{definition}
  
  \newtheorem{definition}{Definition}



\crefname{equation}{equation}{equations}
\Crefname{equation}{Equation}{Equations}
\AtBeginEnvironment{appendices}{\crefalias{section}{appendix}}
\AtBeginEnvironment{appendices}{\crefalias{subsection}{appendix}}
\AtBeginEnvironment{appendices}{\crefalias{subsubsection}{appendix}}
\crefname{appsec}{appendix}{appendices}
\Crefname{appsec}{Appendix}{Appendices}
\Crefname{appendices}{Appendix}{Appendices}
\crefname{appendices}{appendix}{appendices}
\crefname{assumption}{assumption}{assumptions}
\Crefname{assumption}{Assumption}{Assumptions}
\Crefname{lemma}{Lemma}{Lemmata}

\def\id{\mathrm{id}} 

\begin{document}

\title{Tight Samurai Accountant
    }%

\author{Deniz Kattwinkel\thanks{\protect University College London, Department of Economics, \textit{\href{mailto:d.kattwinkel@ucl.ac.uk}{d.kattwinkel@ucl.ac.uk}}} \and Justus Preusser\thanks{\protect Bocconi University, Department of Economics and IGIER, \textit{\href{mailto:justus.preusser@unibocconi.it}{justus.preusser@unibocconi.it}}}}

\date{
  This version: \today%
  \\
}

\maketitle

\begin{abstract}
This note applies tightness (\citet{kattwinkel2025division}) to the setting of \citet[``Samurai Accountant: A Theory of Auditing and Plunder'']{border1987samurai}.
Border and Sobel characterize efficient mechanisms and argue that efficiency entails no loss of optimality.
We characterize tight mechanisms and argue that tightness entails no loss of optimality. 
We show that tight mechanisms form a subset of efficient mechanisms.
Therefore, tightness refines efficiency without loss of optimality. 
By characterizing tight mechanisms, one can replicate the insights from \citet{border1987samurai} and \citet{chander1998general}. 
A novel insight is how and in which order the principal uses different instruments to provide incentives to different agent types.
Further, we describe a procedure for constructing efficient mechanisms in a setting with a continuum of types.
\end{abstract}

\section{Model}\label{sec:model}

\paragraph*{Setup.}
There are a principal and an agent.
The principal has private funds $\tau \geq 0$.
The agent holds a surplus $x \in [\ubar{x}, \bar{x}]$, where $0 \leq \ubar{x} < \bar{x} < \infty$. 
The surplus $x$ is the agent's private information: the agent's \emph{type}.
We do not specify a type distribution as it plays no role.

The agent can make an \emph{advance payment} $y \in [0, x]$ to the principal.
The advance payment is contractible and proves the existence of the advanced portion of the surplus. 

In addition, the principal can acquire conclusive \emph{evidence} about the true surplus, and, thus, make the true surplus contractible.
The agent's advance payment $y$ reveals that the surplus $x$ is at least $y$, and evidence reveals $x$ exactly.
To obtain evidence with probability $a\in [0, 1]$, the principal incurs a cost $c(a)$.
Keeping with \citet{border1987samurai}, we refer to $a$ as the \emph{audit probability}.
The audit probability is contractible, and the principal can attempt to audit only once.
The cost function $c$ is continuous and strictly increasing.\footnote{For linear costs---$c(a)=k\cdot a$ for $k > 0$---, our model exactly replicates \citet{border1987samurai}.} 

The set of feasible transfers depends on the agent's advance payment and whether the principal acquires evidence.
Specifically, if type $x$ advances $y$ and the principal does not audit, then the transfer $t$ from the agent to the principal must be in $[-\tau, y]$; if the principal audits, the transfer $t$ must be in $[-\tau, x]$. 
Given a transfer $t$ and audit probability $a$, the ex-post payoffs of the agent and the principal, respectively, are $x - t$ and $t - c(a)$, respectively.

\paragraph{Mechanisms.} We analyze principal-optimal mechanisms. 
\begin{definition}
A \emph{tax mechanism} is given by a triple $(a, r_{P}, r_{\emptyset})$ of functions and plays out as follows:
\begin{enumerate}
    \item The agent makes an advance payment $y\in [0, \bar{x}]$ to the principal.
    \item The principal audits with probability $a(y)$. If $y < \ubar{x}$, the principal audits with certainty.
    \item \begin{enumerate}
    \item If there is evidence showing that the agent's advance payment is different from the full surplus, the principal seizes the full surplus.
    \item Otherwise, the principal seizes the advance payment and pays the following refund to the agent: 
    \begin{enumerate}
        \item $r_P(y)$ if the principal audited the agent, for a total transfer $y - r_{P}(y)$;
        \item $r_\emptyset(y)$ if the principal did not audit, for a total transfer $y - r_{\emptyset}(y)$.
    \end{enumerate}
    \end{enumerate}
\end{enumerate}

A tax mechanism is \emph{feasible} if for all $y\in [0, \bar{x}]$ the refunds $r_{P}(y)$ and $r_{\emptyset}(y)$ are in $[0, y+\tau]$ and $a(y)$ is a probability.
A tax mechanism is \emph{incentive compatible (IC)} if each type $x$ of the agent has a best response to advance the full surplus ($y=x$).
\end{definition}

A version of the Revelation Principle and basic optimality considerations show that feasible IC tax mechanisms suffice for maximizing the principal's profit (see  \citet{kattwinkel2025division} for a complete argument in a nearby setting).
Henceforth, these are simply called \emph{mechanisms}.
    
\paragraph*{Incentives and profit.}
Fix a mechanism $m$ and type $x$.
The principal's revenue $R_{m}(x)$ and profit $\Pi_{m}(x)$ when facing type $x$ are given by
\begin{align*}
    &R_{m}(x) = x - (a(x) r_{P}(x) + (1 - a(x)) r_{\emptyset}(x))
    \\
    &\Pi_{m}(x) = R_{m}(x) - c(a(x)).
\end{align*}

When advancing $x$ truthfully, type $x$'s utility is given by
\begin{equation*}
    U_{m}(x) = a(x) r_{P}(x) + (1 - a(x))r_{\emptyset}(x) = x - R_{m}(x).
\end{equation*}
When, instead, advancing a strictly smaller amount $y\in [\ubar{x}, x)$, type $x$'s utility is given by $x - (a(y)x + (1 - a(y))(y - r_{\emptyset}(y))$.
Let $\lambda_{m}(x)$ denote type $x$'s lowest deviation loss:
\begin{equation*}
    \lambda_{m}(x) = \inf_{y\in [\ubar{x}, x]} a(y) x + (1 - a(y)) (y - r_{\emptyset}(y)).
\end{equation*}
Incentive compatibility is equivalent to $U_{m}(x) \geq x - \lambda_{m}(x)$ for all $x$; equivalently, $\lambda_{m}(x) \geq R_{m}(x)$.
Note that we may ignore advance payments strictly below $\ubar{x}$ since these payments are audited with certainty and, hence, do not constitute profitable deviations.

\section{Efficiency and tightness}
The notion of efficiency follows \citet{border1987samurai}. In words, a mechanism is efficient if the principal cannot obtain a type-by-type higher revenue while auditing each type with a lower probability.
\begin{definition}[Efficiency]
    A mechanism $m^{\ast}$ is \emph{more efficient than} a mechanism $m$ if $R_{m}\leq R_{m^{\ast}}$ and $a^{\ast} \leq a$.\footnote{For real-valued functions $g$ and $g^{\ast}$, we write $g \leq g^{\ast}$ to mean that $g(x) \leq g^{\ast}(x)$ holds for all $x$. Similarly, $(g, h)\leq (g^{\ast}, h^{\ast})$ means that both $g\leq g^{\ast}$ and $h\leq h^{\ast}$ hold.}
    A mechanism $m^{\ast}$ is \emph{efficient} if $m^{\ast}$ is more efficient than every mechanism $m$ that is more efficient than $m^{\ast}$, i.e. $R_{m^{\ast}}\leq R_{m}$ and $a \leq a^{\ast}$ only if $R_{m} = R_{m^{\ast}}$ and $a = a^{\ast}$.
\end{definition}

We define tightness analogously to \citet{kattwinkel2025division}. In words, a mechanism is tight if the principal cannot obtain a type-by-type higher profit while threatening each type with a higher loss from deviating.
\begin{definition}[Tightness]
    A mechanism $m^{\ast}$ is \emph{tighter than} a mechanism $m$ if $(\Pi_{m}, \lambda_{m}) \leq (\Pi_{m^{\ast}}, \lambda_{m^{\ast}})$.
    A mechanism $m^{\ast}$ is \emph{tight} if $m^{\ast}$ is tighter than every mechanism $m$ that is tighter than $m^{\ast}$, i.e. $(\Pi_{m^{\ast}}, \lambda_{m^{\ast}}) \leq (\Pi_{m}, \lambda_{m})$ only if $(\Pi_{m^{\ast}}, \lambda_{m^{\ast}}) = (\Pi_{m}, \lambda_{m})$.
\end{definition}

\begin{lemma}
    For all mechanisms $m$ there is a tight mechanism that is tighter than $m$.
\end{lemma}
This lemma can be shown using Zorn's Lemma; see \citet{kattwinkel2025division} for a similar argument in their setting.

Below, we also show that every tight mechanism is efficient.
Thus, tightness refines efficiency, without loss for optimality.

\subsection{Characterization}

Let $\Lambda$ denote the set of weakly increasing, weakly concave functions $\lambda\colon [\ubar{x}, \bar{x}]\to [\ubar{x}, \bar{x}]$ such that $\lambda(\ubar{x}) = \ubar{x}$ and $\lambda\leq \id$.
Given $\lambda\in\Lambda$, define $\alpha_{\lambda}$ and $\beta_{\lambda}$ for all $y\in [\ubar{x}, \bar{x}]$ by\footnote{Both suprema are weakly less than $1$ since $\lambda\leq \id$.}
\begin{align}
    \alpha_{\lambda}(y) &= 
    \begin{cases}
        \max\left\lbrace 0, \sup_{x\in (y, \bar{x}]} \frac{\lambda(x) - y}{x - y}\right\rbrace,\quad&\mbox{if }x < \bar{x};
        \\
        0, \quad&\mbox{if }x=\bar{x}.    
    \end{cases}
    \nonumber
    \\
    \beta_{\lambda}(y) &= 
    \begin{cases}
        \max\left\lbrace 0, \sup_{x\in [y, \bar{x}]} \frac{\lambda(x) - \lambda(y)}{x + \tau}\right\rbrace,\quad&\mbox{if }x < \bar{x};
        \\
        0, \quad&\mbox{if }x=\bar{x}.  
    \end{cases}
    \label{eq:alpha_beta}
\end{align}
Both $\alpha_{\lambda}$ and $\beta_{\lambda}$ are weakly decreasing.
\begin{theorem}\label{thm:tight_efficient_characterization}
    If $m$ is tight or efficient, then $\lambda_{m}\in\Lambda$ and $R_{m} = \lambda_{m}$ and $a = \max\lbrace\alpha_{\lambda_{m}}, \beta_{\lambda_{m}}\rbrace$.\footnote{We mean the pointwise maximum, i.e. $a(x) = \max\lbrace\alpha_{\lambda_{m}}(x), \beta_{\lambda_{m}}(x)\rbrace$ for all $x\in[\ubar{x},\bar{x}]$.}
\end{theorem}
The proof also describes the optimal refunds. Namely (after possibly re-defining the refunds without affecting revenue, the agent's utility, or the audit probability): (i) when the audit probability is $\alpha_{\lambda_{m}}$, then the no-evidence refund $r_{\emptyset}$ is $0$ and the audit refund $r_{P}$ is set so that $R_{m} = \lambda_{m}$ holds; else, (ii) the audit refund is maxed out---$r_{P}(y) = y + \tau$---and the no-evidence refund $r_{\emptyset}$ is set so that $R_{m} = \lambda_{m}$ holds.
The next lemma shows that case (i) applies exactly for all sufficiently small types.
\begin{lemma}\label{lemma:alpha_scp}
    For all $\lambda\in\Lambda$, the difference $\alpha_{\lambda} - \beta_{\lambda}$ single-crosses $0$ from above, and $\alpha_{\lambda}(\ubar{x}) \geq \beta_{\lambda}(\ubar{x})$ holds. 
\end{lemma}

The next theorem provides a constructive existence proof of efficient mechanisms.
\begin{theorem}\label{thm:suff_conditions_for_efficiency}
    Let $\lambda\in\Lambda$ and $a = \max\lbrace\alpha_{\lambda}, \beta_{\lambda}\rbrace$.
    \begin{enumerate}
        \item If $(r_{P}, r_{\emptyset})$ are refunds such that $m = (a, r_{P}, r_{\emptyset})$ is a mechanism\footnote{Recall the convention that being a mechanism entails both feasibility and IC.} with revenue given by $R_{m} = \lambda$, then $m$ is efficient.
        \item There exist refunds $(r_{P}, r_{\emptyset})$ such that $m = (a, r_{P}, r_{\emptyset})$ is a mechanism with revenue given by $R_{m} = \lambda$.
    \end{enumerate}
\end{theorem}
The proof constructs the refunds explicitly (see the proof of \Cref{lemma:optimal_refunds}).

An important class of mechanisms are mechanisms $m$ with non-random audits, i.e. $a(y)\in\lbrace 0, 1\rbrace$ for all $y$.
Inspecting the definitions of the audit probabilities $\alpha_{\lambda_{m}}$ and $\beta_{\lambda_{m}}$ in \eqref{eq:alpha_beta}, one can check that a tight or efficient mechanism has non-random audits if and only if there is $y_{0}\in [\ubar{x}, \bar{x}]$ such that the loss function $\lambda_{m}$ is given by $\lambda_{m}(y) = \min\lbrace y, y_{0}\rbrace$ for all $y\in [\ubar{x}, \bar{x}]$; in this case, the principal audits all advance payments strictly below $y_{0}$ and else does not audit, and the revenue $R_{m}$ ($=\lambda_{m}$) is the revenue of a debt contract.
Conversely, if $\lambda\in\Lambda$ is not of the form $\lambda(y) = \min\lbrace y, y_{0}\rbrace$ for some $y_{0}$ and all $y$, then \Cref{thm:suff_conditions_for_efficiency} yields an efficient mechanism with random audits.

\Cref{thm:tight_efficient_characterization} and claim (1) of \Cref{thm:suff_conditions_for_efficiency} imply:
\begin{corollary}
    Every tight mechanism is efficient.
\end{corollary}

\subsection{Proofs}

Let $\Lambda_{0}$ be the set of functions $\lambda\colon [\ubar{x}, \bar{x}]\to\mathbb{R}$ such that $\lambda(x) \in [-\tau, x]$ for all $x$.
For every mechanism $m$, the loss function $\lambda_{m}$ is in $\Lambda_{0}$.

For the analysis, it is useful to ask whether, for given function $\lambda\in\Lambda_{0}$ and audit probabilities $a\colon[\ubar{x}, \bar{x}]\to [0, 1]$, there are feasible refunds $(r_{P}, r_{\emptyset})$ such that revenue equals $\lambda$ and the agent's loss function lies type-by-type above $\lambda$ (which implies IC).
The next lemma shows that such refunds exist if and only if $(\lambda, a)$ satisfies a system of inequalities.
Along the way, we derive the optimal such refunds.
Intuitively, the no-evidence refund $r_{\emptyset}(y)$ should be non-zero only if the audit refund $r_{P}(y)$ equals its upper bound $y + \tau$; this is because $r_{\emptyset}(y)$ attracts other types to deviate to $x$.

Given $\lambda\in\Lambda_{0}$ and $a\colon[\ubar{x}, \bar{x}]\to [0, 1]$, consider the following system of inequalities: $x, y\in [\ubar{x}, \bar{x}]$ such that $y\leq x$,
\begin{equation}\label{eq:optimal_refund_IC}
\lambda(x) \leq a(y) x + \min\left\lbrace(1 - a(y))y, \lambda(y) + a(y)\tau\right\rbrace
.
\end{equation}
Roughly speaking, there are optimally chosen refunds such that the minimum in \eqref{eq:optimal_refund_IC} is given by $(1 - a(y))y$ if $r_{\emptyset}(y) = 0$, and it is given by  $\lambda(y) + a(y)\tau$ if $r_{P}(y) = y + \tau$.

\begin{lemma}\label{lemma:optimal_refunds}
    If $m$ is a mechanism, then the pair $(\lambda_{m}, a)$ satisfies \eqref{eq:optimal_refund_IC}.
    
    Conversely, if $\lambda\in\Lambda_{0}$ and $a\colon[\ubar{x}, \bar{x}]\to [0, 1]$ satisfy \eqref{eq:optimal_refund_IC}, then there exist $(r_{P}, r_{\emptyset})$ such that $m = (a, r_{P}, r_{\emptyset})$ is a mechanism and $\lambda = R_{m} \leq \lambda_{m}$ holds. 
\end{lemma}

Using \Cref{lemma:optimal_refunds}, we will see that tight or efficient mechanisms can be fully characterized in terms of pairs $(\lambda, a)$ consisting of a revenue $\lambda$ and audit probabilities $a$; the refunds are chosen in the background.

\begin{proof}[Proof of \Cref{lemma:optimal_refunds}]
    Let $m$ be a mechanism. We show \eqref{eq:optimal_refund_IC}.
    Fix $y$.
    Since $m$ is IC for type $y$, it holds $(1 - a(y))(y - r_{\emptyset}(y)) + a(y) (y - r_{P}(y)) \leq \lambda_{m}(y)$.
    Since $r_{P}(y) \leq y + \tau$, it follows $(1 - a(y))(y - r_{\emptyset}(y)) \leq \lambda_{m}(y) + a(y) \tau$.
    Since also $r_{\emptyset}(y) \geq 0 $, also $(1 - a(y))(y - r_{\emptyset}(y)) \leq (1 - a(y))y$.
    Thus, $(1 - a(y))(y - r_{\emptyset}(y)) \leq \min\left\lbrace(1 - a(y))y, \lambda_{m}(y) + a(y)\tau\right\rbrace$.
    Therefore, for all $x$ such that $y\leq x$, IC for type $x$ implies $\lambda_{m}(x) \leq a(y) x + (1 - a(y))(y - r_{\emptyset}(y)) \leq  a(y) x + \min\left\lbrace(1 - a(y))y, \lambda_{m}(y) + a(y)\tau\right\rbrace$.
    Thus, \eqref{eq:optimal_refund_IC}.

    Now let $\lambda\in\Lambda_{0}$ and $a\colon[\ubar{x}, \bar{x}]\to [0, 1]$ satisfy \eqref{eq:optimal_refund_IC}.
    For all $y$, choose $(r_{P}(y), r_{\emptyset}(y))\in [0, y+\tau]^{2}$ satisfying
    \begin{align}
        \label{eq:lemma:optimal_refunds:1}
        (1 - a(y))(y - r_{\emptyset}(y)) &= \min\left\lbrace(1 - a(y))y, \lambda(y) + a(y)\tau\right\rbrace,
        \\
        \label{eq:lemma:optimal_refunds:2}
        a(y) (y - r_{P}(y)) &= \lambda(y) - (1 - a(y)) (y - r_{\emptyset}(y)).
    \end{align}
    The following calculations verify that such $(r_{P}(y), r_{\emptyset}(y))\in [0, y+\tau]^{2}$ exist:
    \begin{itemize}
        \item Consider, $r_{\emptyset}(y)$.
        If $a(y) = 1$, then \eqref{eq:lemma:optimal_refunds:1} holds (for an arbitrary choice of $r_{\emptyset}(y)$) since $\lambda(y) + \tau \geq 0$ (since $\lambda\in\Lambda_{0}$).
        If $a(y) < 1$, choose $r_{\emptyset}(y)$ to satisfy \eqref{eq:lemma:optimal_refunds:1}. Then, $(1 - a(y))(y - r_{\emptyset}^{\ast}(y)) \leq (1 - a(y))y$ implies $r_{\emptyset}(y) \geq 0$.
        Further, since $\lambda\in\Lambda_{0}$, we have $\lambda(y) \geq -\tau$ and, thus, $\min\left\lbrace(1 - a(y))y, \lambda_{m}(y) + a(y)\tau\right\rbrace \geq - (1 - a(y))\tau$.
        In particular $(1 - a(y))(y - r_{\emptyset}(y)) \geq - (1 - a(y))\tau$, which implies $r_{\emptyset}(y) \leq y + \tau$. 
        Thus $r_{\emptyset}(y) \in [0, y + \tau]$.
        \item Consider $r_{P}(y)$.
        By the choice of $r_{\emptyset}(y)$, the right side \eqref{eq:lemma:optimal_refunds:2} is given by $\max\lbrace\lambda(y) - (1 - a(y)) y, - a(y)\tau\rbrace$.
        If $a(y) = 0$, this maximum equals $0$ (since $\lambda\leq\id$), and hence \eqref{eq:lemma:optimal_refunds:2} holds for an arbitrary choice of $r_{P}(y)$.
        If $a(y) > 0$, pick $r_{P}(y)$ to solve \eqref{eq:lemma:optimal_refunds:2}.
        Thus, $a(y) (y - r_{P}(y)) \geq -a(y)\tau$, which implies $r_{P}(y) \leq y + \tau$.
        Further, since $\lambda(y)\leq y$, we have $a(y) (y - r_{P}(y)) \leq \max\lbrace y - (1 - a(y)) y, - a(y)\tau\rbrace = a(y) y$, which implies $r_{P}(y) \geq 0$.
        Thus, $r_{P}(y) \in [0, y+\tau]$.
    \end{itemize}
    By construction, for all $y$ it holds
    \begin{equation*}
        \lambda(y) = y - (a(y)r_{P}(y) + (1 - a(y)) r_{\emptyset}(y)) = R_{m}(y).
    \end{equation*}
    It remains to verify $\lambda\leq \lambda_{m}$.
    Fix arbitrary $y, x$ such that $y\leq x$.
    By assumption, $\lambda(x) \leq a(y) x + \min\lbrace (1 - a(y))y, \lambda(y) + a(y)\tau\rbrace$. By the choice of $r_{\emptyset}(y)$, thus also $\lambda(x) \leq a(y) x + (1 - a(y))(y - r_{\emptyset}(y))$.
    Since $\lambda_{m}(x)$ equals the infimum of the right side across all $y$ such that $y\leq x$, we infer $\lambda(x) \leq \lambda_{m}(x)$.    
\end{proof}

The next lemma will be used to show that in all mechanisms that are tight or efficient the loss function is in $\Lambda$. Recall that $\Lambda$ is the set of weakly increasing, weakly concave functions $\lambda\colon [\ubar{x}, \bar{x}]\to [\ubar{x}, \bar{x}]$ such that $\lambda(\ubar{x}) = \ubar{x}$ and $\lambda\leq\id$.
\begin{lemma}\label{lemma:regular_lambda}
    Let $(\lambda, a)$ satisfy \eqref{eq:optimal_refund_IC}.
    Define $\lambda^{+}(x) = \max\lbrace\ubar{x}, \sup_{y\in [\ubar{x}, x]} \lambda(y)\rbrace$ and
    \begin{equation}\label{lemma:virtual_loss_definition}
        \lambda^{\ast}(x) = \inf_{y\in [\ubar{x}, \bar{x}]} a(y) x + \min\left\lbrace(1 - a(y))y, \lambda^{+}(y) + a(y)\tau\right\rbrace.
    \end{equation}
    Then $\lambda^{\ast}\in\Lambda$ and $\lambda \leq \lambda^{\ast}$ and the pair $(\lambda^{\ast}, a)$ satisfies \eqref{eq:optimal_refund_IC}.
\end{lemma}
Note that the infimum in \eqref{lemma:virtual_loss_definition} is taken over all types, in contrast to the definition of a mechanism's loss function.
Loosely speaking, the lemma thus shows that when the refunds $(r_{P}, r_{\emptyset})$ are chosen optimally (as constructed in the proof of \Cref{lemma:optimal_refunds}), then only downward deviations are relevant.

In a tight or efficient mechanism $m$, we will see in \Cref{thm:tight_efficient_characterization} that the loss function $\lambda_{m}$ is in $\Lambda$; increasingness of $\lambda_{m}$ then says that higher types lose when their deviations are detected; concavity of $\lambda_{m}$ stems from the fact that the agent deviates to minimize loss; $\lambda_{m}(\ubar{x}) = \ubar{x}$ means the lowest type expects to lose everything when deviating; and $\lambda_{m}\leq \id$ simply means that the agent can never lose more than the surplus.

\begin{proof}[Proof of \Cref{lemma:regular_lambda}]
    First, note that $\lambda\leq \lambda^{+}\leq\id$ and $\lambda^{+}(\ubar{x}) = \ubar{x}$ hold, and that $\lambda^{+}$ is weakly increasing.
    
    To verify $\lambda^{\ast}\in\Lambda$, note that $\lambda^{\ast}$ is weakly increasing and weakly concave as $\lambda^{\ast}$ is the pointwise infimum of weakly increasing, affine functions.
    Finally, $\lambda^{\ast}(\ubar{x}) = \ubar{x}$ and $\lambda^{\ast}\leq\id$ follow from inspecting \eqref{lemma:virtual_loss_definition} and using $\lambda^{+}\leq\id$ and $\lambda^{+}(\ubar{x}) = \ubar{x}$.
    
    In an intermediate step, we show $\lambda^{+}(x) \leq \lambda^{\ast}(x)$ for all $x$.
    By inspection, $\ubar{x}\leq\lambda^{\ast}(x)$.
    Thus, we show $\sup_{x^{\prime}\in [\ubar{x}, x]} \lambda(x^{\prime}) \leq \lambda^{\ast}(x)$; i.e. for all $x^{\prime} \in [\ubar{x}, x]$ and all $y$,
    \begin{equation*}
        \lambda(x^{\prime}) \leq a(y) x + \min\left\lbrace(1  - a(y))y, \lambda^{+}(y) + a(y)\tau\right\rbrace.
    \end{equation*}
    First, if $x^{\prime} \leq y$, then since $\lambda^{+}$ is weakly increasing and $\lambda(x^{\prime}) \leq \lambda^{+}(x^{\prime}) \leq x^{\prime}$, it holds 
    \begin{align*}
        \lambda(x^{\prime}) \leq \lambda^{+}(x^{\prime}) &\leq a(y) x^{\prime} + \min\left\lbrace(1  - a(y))x^{\prime}, \lambda^{+}(x^{\prime})\right\rbrace
        \\
        &\leq a(y) x + \min\left\lbrace(1  - a(y))y, \lambda^{+}(y) + a(y)\tau\right\rbrace,
    \end{align*}    
    as desired.
    Second, if $y \leq x^{\prime}$, then \eqref{eq:optimal_refund_IC} and $\lambda(y) \leq \lambda^{+}(y)$ imply 
    \begin{align*}
        \lambda(x^{\prime}) &\leq a(y) x^{\prime} + \min\left\lbrace(1  - a(y))y, \lambda(y) + a(y)\tau\right\rbrace 
        \\ & \leq a(y) x + \min\left\lbrace(1  - a(y))y, \lambda^{+}(y) + a(y)\tau\right\rbrace,
    \end{align*}
    Thus, $\lambda^{+}(x) \leq \lambda^{\ast}(x)$ for all $x$.

    It remains to verify $\lambda \leq \lambda^{\ast}$ and that the pair $(\lambda^{\ast}, a)$ satisfies \eqref{eq:optimal_refund_IC}.
    First, $\lambda\leq\lambda^{\ast}$ follows since $\lambda\leq \lambda^{+}$ (by definition of $\lambda^{+}$) and $\lambda^{+} \leq \lambda^{\ast}$ (proven earlier).
    To show that $(\lambda^{\ast}, a)$ satisfies \eqref{eq:optimal_refund_IC}, fix $x, y$ such that $y\leq x$.
    Then (the first inequality follows from the definition of $\lambda^{\ast}$; the second from $\lambda^{+}\leq\lambda^{\ast}$)
    \begin{align*}
        \lambda^{\ast}(x) \leq & a(y) x + \min\left\lbrace(1 - a(y))y, \lambda^{+}(y) + a(y)\tau\right\rbrace
        \\
        \leq & a(y) x + \min\left\lbrace(1 - a(y))y, \lambda^{\ast}(y) + a(y)\tau\right\rbrace,
    \end{align*}
    as desired.
\end{proof}

Intuitively, the audit probability should be chosen as small as possible to save costs.
Fixing $\lambda$, the lowest possible value audit probability that is feasible for \eqref{eq:optimal_refund_IC} is given by the probabilities $\alpha_{\lambda}$ and $\beta_{\lambda}$ from \eqref{eq:alpha_beta}.
The next lemma combines this intuition and the previous two lemmata to show that, by possibly passing to a more efficient and tighter mechanism, we may focus on revenues $\lambda^{\ast}$ in $\Lambda$ and audit probabilities given by $a^{\ast} = \max\lbrace \alpha_{\lambda^{\ast}}, \beta_{\lambda^{\ast}}\rbrace$.
\begin{lemma}\label{lemma:tightened}
    Let $m = (a, r_{P}, r_{\emptyset})$ be a mechanism.
    There exists $\lambda^{\ast}\in\Lambda$ and a mechanism $m^{\ast} = (a^{\ast}, r_{P}^{\ast}, r_{\emptyset}^{\ast})$ such that
    \begin{equation*}
        a^{\ast} = \max\lbrace \alpha_{\lambda^{\ast}}, \beta_{\lambda^{\ast}}\rbrace \leq a
        \quad\mbox{and}\quad
        \lambda_{m} \leq \lambda^{\ast} = R_{m^{\ast}} \leq \lambda_{m^{\ast}}
    \end{equation*}
    In particular, $m^{\ast}$ is more efficient and tighter than $m$.
\end{lemma}

\begin{proof}[Proof of \Cref{lemma:tightened}]
    \Cref{lemma:optimal_refunds} implies that $(\lambda_{m}, a)$ satisfies \eqref{eq:optimal_refund_IC}.
    By \Cref{lemma:regular_lambda}, there exists $\lambda^{\ast}\in\Lambda$ such that $\lambda_{m}\leq\lambda^{\ast}$ and such that $(\lambda^{\ast}, a)$ also satisfies \eqref{eq:optimal_refund_IC}.
    Define $a^{\ast} = \max\lbrace \alpha_{\lambda^{\ast}}, \beta_{\lambda^{\ast}}\rbrace$.
    Fixing $y$, consider the set of $\tilde{a}\in [0, 1]$ such that 
    \begin{equation}\label{eq:auxiliary_problem_IC:restated}
        \forall x\in [y, \bar{x}],\quad
        \lambda^{\ast}(x) \leq \tilde{a} x + \min\left\lbrace(1 - \tilde{a})y, \lambda^{\ast}(y) + \tilde{a}\tau\right\rbrace.
    \end{equation}
    As just argued, $(\lambda^{\ast}, a)$ satisfies \eqref{eq:optimal_refund_IC}, meaning $a(y)$ satisfies \eqref{eq:auxiliary_problem_IC:restated} for all $y$.
    By inspection, the smallest value $\tilde{a}\in [0, 1]$ satisfying \eqref{eq:auxiliary_problem_IC:restated} equals $a^{\ast}(y) = \max\lbrace\alpha_{\lambda_{m}}(y), \beta_{\lambda_{m}}(y)\rbrace$.
    Since $a^{\ast}(y)$ satisfies \eqref{eq:auxiliary_problem_IC:restated} for all $y$, moreover, \Cref{lemma:optimal_refunds} implies that there exist refunds $(r_{P}^{\ast}, r_{\emptyset}^{\ast})$ such that $m^{\ast} = (a^{\ast}, r_{P}^{\ast}, r_{\emptyset}^{\ast})$ is a mechanism with $\lambda_{m}\leq \lambda^{\ast} = R_{m^{\ast}} \leq \lambda_{m^{\ast}}$.
    As just argued, moreover, $a^{\ast}\leq a$.
\end{proof}

\begin{proof}[Proof of \Cref{thm:tight_efficient_characterization}]
    Let $m$ be tight or efficient.
    Invoking \Cref{lemma:tightened}, there exists $\lambda^{\ast}\in\Lambda$ and a mechanism $m^{\ast} = (a^{\ast}, r_{P}^{\ast}, r_{\emptyset}^{\ast})$ such that
    \begin{equation*}
        a^{\ast} = \max\lbrace \alpha_{\lambda^{\ast}}, \beta_{\lambda^{\ast}}\rbrace \leq a
        \quad\mbox{and}\quad
        \lambda_{m} \leq \lambda^{\ast} = R_{m^{\ast}} \leq \lambda_{m^{\ast}}
    \end{equation*}
    Since $m$ is IC, also $R_{m}\leq \lambda_{m}$.
    From $R_{m} \leq R_{m^{\ast}}$ and $a^{\ast}\leq a$, we also infer $\Pi_{m}\leq \Pi_{m^{\ast}}$, and, since the audit probability strictly decreases profits, we have $\Pi_{m}\neq \Pi_{m^{\ast}}$ if $a^{\ast} \neq a$.
    
    Now, if $m$ is efficient, then $R_{m} \leq R_{m^{\ast}}$ and $a^{\ast}\leq a$ imply $R_{m} = R_{m^{\ast}}$ and $a^{\ast}= a$. Hence, also $\lambda_{m} = R_{m}$ and $\lambda_{m} \in \Lambda$ and $a = \max\lbrace \alpha_{\lambda_{m}}, \beta_{\lambda_{m}}\rbrace$, as desired.
    
    If $m$ is tight, then $\Pi_{m}\leq \Pi_{m^{\ast}}$ and $\lambda_{m} \leq \lambda_{m^{\ast}}$ imply $\Pi_{m}= \Pi_{m^{\ast}}$ and $\lambda_{m} = \lambda_{m^{\ast}}$. Hence, also $a= a^{\ast}$ (since else $\Pi_{m} \neq \Pi_{m^{\ast}}$). From $\lambda_{m} = \lambda_{m^{\ast}}$, we also find $\lambda_{m} = R_{m}$ and $\lambda_{m} \in \Lambda$ and $a = \max\lbrace \alpha_{\lambda_{m}}, \beta_{\lambda_{m}}\rbrace$, as desired.
\end{proof}

\begin{proof}[Proof of \Cref{lemma:alpha_scp}]
    Fix $y$.
    The probability $\alpha_{\lambda_{m}}(y)$ is the smallest probability $\tilde{a}$ such that $\lambda_{m}(x) \leq \tilde{a} x + (1 - \tilde{a})y$ for all $x\in [y, \bar{x}]$, whereas $\beta_{\lambda_{m}}(y)$ is the smallest probability $\tilde{a}$ such that $\lambda_{m}(x) \leq \tilde{a} x + \lambda_{m}(y) + \tilde{a}\tau$ for all $x\in [y, \bar{x}]$.
    Thus, $\alpha_{\lambda_{m}}(y) \leq \beta_{\lambda_{m}}(y)$ if and only if $(1 - \alpha_{\lambda_{m}}(y))y \geq \lambda_{m}(y) + \alpha_{\lambda_{m}}(y)\tau$.
    Equivalently, $1 - \alpha_{\lambda_{m}}(y) (1 + \frac{\tau}{y}) \geq \frac{\lambda_{m}(y)}{y}$.
    Notice that $ \alpha_{\lambda_{m}}$ is weakly decreasing, meaning the left side of this inequality is weakly increasing in $y$.
    However, $\lambda_{m}(y) / y$ is weakly decreasing since $\lambda_{m}\in\Lambda$.
    Thus, $\alpha_{\lambda_{m}}(y) - \beta_{\lambda_{m}}(y)$ single-crosses $0$ from above.
    Finally, the inequality $\alpha_{\lambda_{m}}(\ubar{x}) \geq \beta_{\lambda_{m}}(\ubar{x})$ follows from this argument and from observing that $\lambda_{m}(\ubar{x}) = \ubar{x}$ holds.
\end{proof}

The following lemma is the key step towards \Cref{thm:suff_conditions_for_efficiency}.
\begin{lemma}\label{lemma:almost_efficient}
    Let $\lambda, \lambda^{\ast}\in\Lambda$, let $a = \max\lbrace \alpha_{\lambda}, \beta_{\lambda}\rbrace$, and let $a^{\ast} = \max\lbrace\alpha_{\lambda^{\ast}}, \beta_{\lambda^{\ast}}\rbrace$.
    If $\lambda\leq\lambda^{\ast}$ and $a^{\ast}\leq a$, then $\lambda=\lambda^{\ast}$.
\end{lemma}

\begin{proof}[Proof of \Cref{lemma:almost_efficient}]
    For later reference, we record that for all $y$, it holds
    \begin{equation}\label{eq:lemma:almost_efficient:IC}
        \forall x\in [y, \bar{x}],\quad \lambda^{\ast}(x) \leq a^{\ast}(y) + \min\lbrace (1 - a^{\ast}(y))y, \lambda^{\ast}(y) + a^{\ast}(y)\tau\rbrace.
    \end{equation}

    Let $\ubar{y} = \max\lbrace y\colon \lambda(y) = y\rbrace$ and $\bar{y} = \min\lbrace y\colon \lambda(y) = \max \lambda\rbrace$.
    Necessarily $\ubar{y}\leq\bar{y}$ since $\lambda\in\Lambda$.

    We distinguish two cases.
    First, suppose $\ubar{y} = \bar{y}$, so that $\lambda$ is given by $\lambda(y) = \min\lbrace y, \ubar{y}\rbrace$ for all $y\in [\ubar{x}, \bar{x}]$.
    For $y\in [\ubar{x}, \ubar{y}]$, we thus have $\lambda^{\ast}(y) \leq y = \lambda(y)$, implying $\lambda^{\ast}(y) = \lambda(y)$.
    For $y\in [\ubar{y}, \bar{x}]$, and using that $\lambda$ is constant on $[\ubar{y}, \bar{x}]$ one may verify from the definitions of $\alpha_{\lambda}$ and $\beta_{\lambda}$ that $\alpha_{\lambda}(y) = \beta_{\lambda}(y) = 0$ hold.
    Thus, $a = 0$ on $[\ubar{y}, \bar{x}]$ and, hence, also $a^{\ast} = 0$ on $[\ubar{y}, \bar{x}]$.
    Inspecting \eqref{eq:lemma:almost_efficient:IC} and using $a^{\ast}(\ubar{y}) = 0$, it follows $\lambda^{\ast}(x) \leq \ubar{y}$ for all $x\in [\ubar{y}, \bar{x}]$.
    But $\ubar{y} = \lambda(x)$ for all $x\in [\ubar{y}, \bar{x}]$.
    Thus, also $\lambda^{\ast} = \lambda$ on $[\ubar{y}, \bar{x}]$.
    We conclude $\lambda^{\ast} = \lambda$, as promised.

    In what follows, assume $\ubar{y} < \bar{y}$.

    For all $y$, let $\phi(y) = \min\lbrace(1 - a(y))y, \lambda(y) + a(y)\tau\rbrace$ and let $\hat{X}(y)$ be the set of $x\in [y, \bar{x}]$ such that
    \begin{equation*}
        \lambda(x) = a(y) x + \phi(y).
    \end{equation*}
    Following \citet{kattwinkel2025division}, one can show that (1) $\hat{X}(y)$ is non-empty for all $y\in [\ubar{x}, \bar{x}]$; (2) for every $y\in (\ubar{y}, \bar{y})$ it holds $y < \min\hat{X}(y)$; (3) for every $y\in[\ubar{x}, \bar{x}]$ the image $\hat{X}([\ubar{x}, y])$ is the interval $[\ubar{y}, \max\hat{X}(y)]$.

    We immediately have $\lambda = \lambda^{\ast}$ on $[\ubar{x}, \ubar{y}]$ since $\lambda(y) = y \geq \lambda^{\ast}(y)$ for all $y\in[\ubar{x}, \ubar{y}]$.

    According to \Cref{lemma:alpha_scp}, the difference $\alpha_{\lambda} - \beta_{\lambda}$ single-crosses $0$ from above and $\alpha_{\lambda}(\ubar{x}) \geq \beta_{\lambda}(\ubar{x})$ holds.
    Let $y_{0}$ be the supremum of the types $y$ such that $\alpha_{\lambda}(y) \geq \beta_{\lambda}(y)$.
    Rearranging and using $a(y) = \max\lbrace \alpha_{\lambda}(y), \beta_{\lambda}(y)\rbrace$, one may verify that $\phi(y) = (1 - a(y))y$ holds for $y < y_{0}$, and $\phi(y) = \lambda(y) + a(y)\tau$ for $y > y_{0}$.

    Using $\ubar{y} < \bar{y}$, one may verify that also $\ubar{y} < y_{0}$ holds.
    It suffices to show $\alpha_{\lambda}(\ubar{y}) > \beta_{\lambda}(y)$.
    As in the proof of \Cref{lemma:alpha_scp}, we have $\alpha_{\lambda}(\ubar{y}) > \beta_{\lambda}(y)$ if and only if $1 - \alpha_{\lambda}(\ubar{y})(1 + \tau/\ubar{y}) < \lambda(\ubar{y}) / \ubar{y}$ holds. The right side equals $1$, by definition of $\ubar{y}$. Since $\ubar{y} < \bar{y}$, the function $\lambda$ is not constant on $[\ubar{y}, \bar{x}]$. In particular, there exists $x > \ubar{y}$ such that $\lambda(x) > \lambda(\ubar{y}) = \ubar{y}$. Now inspecting the definition of $\alpha_{\lambda}$, we infer $\alpha_{\lambda}(\ubar{y}) > 0$. Thus, $1 - \alpha_{\lambda}(\ubar{y})(1 + \tau/\ubar{y}) < \lambda(\ubar{y}) / \ubar{y}$.
    
    We next argue $\lambda = \lambda^{\ast}$ on $(\ubar{y}, y_{0})$.
    Take $x \in (\ubar{y}, y_{0})$.
    By the properties of $\hat{X}$, there is $y\leq x$ such that $x\in\hat{X}(y)$,\footnote{Specifically, the properties imply that the image $\hat{X}([\ubar{x}, x])$ is the integral $[\ubar{x}, \max\hat{X}(x)]$. Since $\hat{X}$ always maps above the identity, $x\leq\max\hat{X}(x)$, and so also $x$ is in the image. Thus there is $y\in[\ubar{x}, x]$ such that $x\in\hat{X}(y)$.} meaning $\lambda(x) = a(y) x + \phi(y)$.
    Since $y< y_{0}$, it holds $\phi(y) = (1 - a(y))y$.
    Since $a \geq a^{\ast}$, also $\lambda(x) \geq a^{\ast}(y) x + (1- a^{\ast}(y))y$.
    But, in view of $y\leq x$ and \eqref{eq:lemma:almost_efficient:IC}, also $\lambda^{\ast}(x) \leq a^{\ast}(y) x + (1- a^{\ast}(y))y$.
    In particular, $\lambda^{\ast}(x) \leq \lambda(x)$.
    Since $\lambda^{\ast}\geq\lambda$ holds by assumption, we conclude $\lambda^{\ast}(x) = \lambda(x)$.

    Since $\lambda$ and $\lambda^{\ast}$ are continuous, we infer that $\lambda$ and $\lambda^{\ast}$ agree on $[\ubar{x}, y_{0}]$.

    Now let $z = \max\lbrace y \in [y_{0}, \bar{y}]\colon \forall y^{\prime}\in[\ubar{x}, y],\, \lambda(y^{\prime}) = \lambda^{\ast}(y^{\prime})\rbrace$.\footnote{By the previous paragraph, the set $\lbrace y \in [y_{0}, \bar{y}]\colon \forall y^{\prime}\in[\ubar{x}, y],\quad \lambda(y^{\prime}) = \lambda^{\ast}(y^{\prime})\rbrace$ is non-empty. The set is compact since $\lambda$ and $\lambda^{\ast}$ are continuous (by definition of $\Lambda$). Thus, $z$ is well-defined.}
    We show $z = \bar{y}$.
    If $y_{0} = \bar{y}$, we are done.
    Thus, let $y_{0} < \bar{y}$.
    Take $y \in [y_{0}, \bar{y})$ and suppose $\lambda(y^{\prime}) = \lambda^{\ast}(y^{\prime})$ for all $y^{\prime} \leq y$.
    We show that $y$ cannot equal $z$.
    Note $y\in (\ubar{y}, \bar{y})$ since $\ubar{y} < y_{0} <\bar{y}$.
    By the properties of $\hat{X}$, it holds $\max\hat{X}(y) > y$.
    Take arbitrary $x\in [y, \max\hat{X}(y)]$.
    Since the image $\hat{X}([\ubar{x}, y])$ equals $[y, \max\hat{X}(y)]$, there exists $y^{\prime} \leq y$ such that $x\in\hat{X}(y^{\prime})$.
    By the assumed properties of $y$, thus $\lambda^{\ast}(y^{\prime}) = \lambda(y^{\prime})$.
    Since $a(y^{\prime})\geq a^{\ast}(y^{\prime})$ and $\lambda^{\ast}(y^{\prime}) = \lambda(y^{\prime})$, we find $\lambda(x) = a(y^{\prime}) x + \lambda(y^{\prime}) + a(y^{\prime})\tau \geq a^{\ast}(y^{\prime}) x + \lambda^{\ast}(y^{\prime}) + a^{\ast}(y^{\prime})\tau$.
    In view of $y^{\prime}\leq x$ and \eqref{eq:lemma:almost_efficient:IC}, also $\lambda^{\ast}(x) \leq a^{\ast}(y^{\prime}) x + \lambda^{\ast}(y^{\prime}) + a^{\ast}(y^{\prime})\tau$.
    Thus, $\lambda \geq \lambda^{\ast}$ and so we conclude $\lambda(x) = \lambda^{\ast}(x)$.
    Since $x\in [\ubar{y}, \max\hat{X}(y)]$ was arbitrary and $\max\hat{X}(y) > y$, it follows that $y$ cannot be $z$, i.e. the largest value in $[y_{0}, \bar{y}]$ such that $\lambda(y) = \lambda^{\ast}(y)$ for all $y \leq z$.
    Thus $z = \bar{y}$.

    Finally, we argue $\lambda = \lambda^{\ast}$ on $[\bar{y}, \bar{x}]$.
    From the definition of $\bar{y}$, recall that $\lambda$ is constantly equal to $\max\lambda$ on $[\bar{y}, \bar{x}]$.
    Inspecting the definitions, it follows that $a = \max\lbrace\alpha_{\lambda}, \beta_{\lambda}\rbrace$ is constantly $0$ on $[\bar{y}, \bar{x}]$.
    Thus, also $a^{\ast}$ is constantly $0$ on $[\bar{y}, \bar{x}]$.
    Using \eqref{eq:lemma:almost_efficient:IC}, $m^{\ast}$ is IC, it now follows readily that $\lambda^{\ast}$ is bounded above by $\lambda^{\ast}(\bar{y})$ on $[\bar{y}, \bar{x}]$.
    Since $\lambda$ and $\lambda^{\ast}$ coincide at $\bar{y}$ (by the previous paragraph), we conclude that $\lambda\geq \lambda^{\ast}$ holds on $[\bar{y}, \bar{x}]$.
    Thus $\lambda= \lambda^{\ast}$ on $[\bar{y}, \bar{x}]$.
\end{proof}

\begin{proof}[Proof of \Cref{thm:suff_conditions_for_efficiency}]
    Given $a = \max\lbrace \alpha_{\lambda}, \beta_{\lambda}\rbrace$, the pair $(\lambda, a)$ satisfies \eqref{eq:optimal_refund_IC} of \Cref{lemma:optimal_refunds}. 
    Thus, \Cref{lemma:optimal_refunds} shows that there are refunds $(r_{P}, r_{\emptyset})$ such that $m = (a, r_{P}, r_{\emptyset})$ is a mechanism with revenue given by $R_{m} = \lambda$.
    It remains to show that for an arbitrary pair of such refunds the mechanism is efficient.

    Let $m^{\dagger}$ with audit probability $a^{\dagger}$ be another mechanism such that $R_{m} \leq R_{m^{\dagger}}$ and $a^{\dagger} \leq a$.
    We have to show $R_{m} = R_{m^{\dagger}}$ and $a^{\dagger} = a$.
    Invoking \Cref{lemma:tightened}, find $(a^{\ast}, \lambda^{\ast})$ such that $a^{\ast} = \max\lbrace\alpha_{\lambda^{\ast}}, \beta_{\lambda^{\ast}}\rbrace$ and $\lambda^{\ast}\in\Lambda$ and $a^{\ast}\leq a^{\dagger}$ and $R_{m^{\dagger}} \leq \lambda^{\ast}$.
    Collecting inequalities, also $a^{\ast} \leq a^{\dagger}\leq a$ and $\lambda = R_{m} \leq R_{m^{\dagger}} \leq \lambda^{\ast}$.
    We now invoke \Cref{lemma:almost_efficient} to infer $\lambda = \lambda^{\ast}$.
    Thus, also $a = a^{\ast}$ (from the definitions of $a$ and $a^{\ast}$) and $\lambda = R_{m} = R_{m^{\dagger}} = \lambda^{\ast}$ (from the earlier inequality chain).
    Finally, the chain $a^{\ast} \leq a^{\dagger} \leq a$ implies $a^{\dagger} = a$.
\end{proof}

\addcontentsline{toc}{section}{References}
\newrefcontext[sorting=nyt]
\printbibliography

\end{document}